\documentclass[12pt,preprint]{aastex}

\newcommand{\dechms}[4]{$#1^{\rm h}#2^{\rm m}#3\mbox{$^{\rm s}\mskip-7.6mu.\,$}#4$}

\newcommand{\decdms}[4]{$#1^{\circ}#2'#3\mbox{$''\mskip-7.6mu.\,$}#4$}
\newcommand{\msec}[2]{$#1\mbox{$''\mskip-7.6mu.\,$}#2$}
\newcommand{\mmsec}[2]{$#1\mbox{$^s\mskip-7.6mu.\,$}#2$}

\newcommand{\HII}{\mbox{H\,{\sc ii}}}

\newcommand{\Msun}{M$_{\odot}$}
\newcommand{\Rsun}{R$_{\odot}$}

\begin{document}

\title{A Preliminary VLBA Distance to the Core of Ophiuchus, with an Accuracy of 4\%}

\author{Laurent Loinard\footnote{Centro de Radiostronom\'{\i}a y
Astrof\'{\i}sica, Universidad Nacional Aut\'onoma de M\'exico,
Apartado Postal 72--3 (Xangari), 58089 Morelia, Michoac\'an, M\'exico;
l.loinard@astrosmo.unam.mx}, Rosa M.\ Torres$^1$, Amy J.\
Mioduszewski\footnote{National Radio Astronomy Observatory, Array 
Operations Center, 1003 Lopezville Road, Socorro, NM 87801, USA},
and Luis F.\ Rodr\'{\i}guez$^1$}

\begin{abstract}

  The non-thermal 3.6 cm radio continuum emission from the young stars
  S1 and DoAr21 in the core of Ophiuchus, has been observed with the
  Very Long Baseline Array (VLBA) at 6 and 7 epochs, respectively,
  between June 2005 and August 2006. The typical separation between
  successive observations was 2 to 3 months. Thanks to the remarkably
  accurate astrometry delivered by the VLBA, the trajectory described
  by both stars on the plane of the sky could be traced very
  precisely, and modeled as the superposition of their trigonometric
  parallax and a uniform proper motion. The best fits yield distances
  to S1 and DoAr21 of 116.9$^{+7.2}_{-6.4}$ pc and 121.9$^{+5.8}_{-5.3}$ pc,
  respectively.  Combining these results, we estimate the mean
  distance to the Ophiuchus core to be 120.0$^{+4.5}_{-4.2}$ pc, a value
  consistent with several recent indirect determinations, but with a
  significantly improved accuracy of 4\%. Both S1 and DoAr21 happen to
  be members of tight binary systems, but our observations are not
  frequent enough to properly derive the corresponding orbital
  parameters. This could be done with additional data, however, and
  would result in a significantly improved accuracy on the distance
  determination.

\end{abstract}


\section{Introduction} 

Ophiuchus is one of the most active regions of star-formation within a
few hundred parsecs of the Sun (e.g.\ Lada \& Lada 2003). It has
played an important role in the development of our understanding of
star-formation, and remains an important benchmark for this field of
research. Indeed, it has been one of the key targets of the Spitzer
c2d legacy program (Padgett et al.\ 2007); and has been observed in
detail at numerous other wavelengths, including X-rays (Ozawa et al.\
2005, Gagn\'e et al.\ 2004), near-infrared (e.g.\ Haisch et al.\ 2002,
Duch\^ene et al.\ 2004, and references therein), sub-millimeter (Motte
et al.\ 1998, Johnstone et al.\ 2004), and radio (e.g.\ Andr\'e et
al.\ 1987, Leous et al.\ 1991).

The detailed analysis of this wealth of observational data has been
somewhat hampered by the relatively large uncertainty on the distance
to the Ophiuchus complex. Traditionally assumed to be at 165 pc (Chini
1981), it has recently been suggested to be somewhat closer. For
example, de Geus et al.\ (1989) found a mean photometric distance of
125 $\pm$ 25 pc. Knude \& Hog (1998), who examined the reddening of
stars in the direction of Ophiuchus as a function of their Hipparcos
distances, also found a clear extinction jump at 120 pc. Using a
similar method, M.\ Lombardi et al.\ (in prep.) also find a distance
of about 120 pc for the Ophiuchus core. Finally, Mamajek (2007)
identified reflection nebulae within 5$^\circ$ of the center of
Ophiuchus, and obtained the trigonometric parallax of the illuminating
stars from the Hipparcos catalog. From the average of these Hipparcos
parallaxes, he obtains a mean distance to Ophiuchus of 135 $\pm$ 8 pc.

This latter result is based on parallax measurements, but considers a
fairly large area around Ophiuchus. It could, therefore, include
objects unrelated to Ophiuchus itself. The former results are
restricted to regions more concentrated on Ophiuchus, but they are
based on indirect distance determinations. Here, we will present
measurements of the trigonometric parallax of two young stars (S1 and
DoAr21) directly associated with the Ophiuchus core. This will allow
us to estimate directly the distance to this important region of
star-formation.

\section{Observed sources}

The star S1 (of spectral type B4, $M$ $\sim$ 6 \Msun) is among the
brightest red and near-infrared objects in Ophiuchus (Grasdalen et
al.\ 1973). It is also the brightest far-infrared member of the
cluster (Fazio et al.\ 1976), a very bright X-ray source (ROX 14
--Montmerle et al.\ 1983), and the brightest steady radio stellar
object in Ophiuchus\footnote{DoAr21 --as shown by Feigelson \&
  Montmerle (1985), and as we shall confirm below-- can occasionally
  become brighter than S1.}  (Leous et al.\ 1991). S1 is fairly
heavily obscured ($A_V$ $\sim$ 10), and there is clear evidence for an
interaction between S1 and the dense gas associated with Oph A, and
traced by DCO$^+$ emission (Loren et al.\ 1990).  Moreover, the age of
the \HII\ region excited by S1 is estimated to be about 5,000 yr
(Andr\'e et al.\ 1988). All this demonstrates that S1 can safely be
assumed to be a member of the Ophiuchus core.

DoAr21 (Dolidze-Arakelyan 21) is a somewhat less massive star ($\sim$
2.2 \Msun) of spectral type K1 (E.\ Jensen et al., in prep.). Like S1,
it is fairly obscured ($A_V$ $\sim$ 6--7), and probably younger than
10$^6$ yr. It is associated with a bright X-ray source (ROX 8
--Montmerle et al.\ 1983), and with a strongly variable radio source
(Feigelson \& Montmerle 1985). Although it has long been classified as
a naked T Tauri star (e.g.\ Andr\'e et al.\ 1990), it was recently
found to show a substantial infrared excess at 25 $\mu$m (Jensen et
al.\ ibid) suggestive of a circumstellar disk. Given its youth, and
location in the Ophiuchus core, DoAr21 is almost certainly also a {\it
bona fide} member of the Ophiuchus complex.

As mentioned above, both S1 and DoAr21 are fairly strong radio
sources.  Indeed, both have been detected at 6 cm in previous Very
Long Baseline Interferometry experiments: S1 with a flux density of
6--9 mJy (Andr\'e et al.\ 1991), and DoAr21 with a flux density of
nearly 10 mJy (Phillips et al.\ 1991).

\section{Observations}

In this paper, we will make use of two series of continuum 3.6 cm
(8.42 GHz) observations obtained with the VLBA. Six observations of S1
were collected between June 2005 and August 2006, and seven
observations of DoAr21 were obtained between September 2005 and August
2006 (See Tab.\ 1 for details). Each observation consisted of series
of cycles with two minutes spent on source, and one minute spent on
the phase-referencing quasar J1625--2527, located 1$^\circ$ south of
both targets. J1625--2527 is a very compact extragalactic source whose
absolute position ($\alpha_{J2000.0}$ =
\dechms{16}{25}{46}{8916},$\delta_{J2000.0}$ =
\decdms{-25}{27}{38}{327}) is known to better than 0.5 milli-arcsecond
(mas --Beasley et al.\ 2002). The data were edited and calibrated
using the Astronomical Image Processing System (AIPS --Greisen
2003). The basic data reduction followed the standard VLBA procedures
for phase-referenced observations, and was described in detail in
Loinard et al.\ (2007). Since the density of compact quasars known
around Ophiuchus at the time of our observations was insufficient, we
could not apply the multi-source calibration described in Torres et
al.\ (2007).

Because of the significant overheads that were necessary to properly
calibrate the data, only about 2 of the 4 hours of telescope time
allocated to each of our observations were actually spent on
source. Once calibrated, the visibilities were imaged with a pixel
size of 50 $\mu$as after weights intermediate between natural and
uniform (ROBUST = 0 in AIPS) were applied. This resulted in typical
r.m.s.\ noise levels of 0.1 to 0.3 mJy depending on the weather
conditions and source strength (Tab.\ 1). Both S1 and DoAr21 were
detected with a signal to noise better than 7 at each epoch (Tab.\ 1).

\section{Results, discussions and conclusions}

\subsection{Properties of S1}

The mean 3.6 cm flux of S1 in our data is 4.8 mJy, and the dispersion
about that mean is 1.2 mJy (see Fig.\ 1). This shows that S1 is
variable at the level of about 25\% on timescales of months to
years. This modest level of variability is certainly not unexpected
for a non-thermal source associated with an active stellar
magnetosphere (Feigelson \& Montmerle 1999). As mentioned earlier,
Andr\'e et al.\ (1991) reported a VLBI detection of S1 at 6 cm. They
found --among many other things-- that the source was somewhat
resolved in their observations, with a full width at half maximum
extension of about 1.7 mas. The radio emission associated with S1 is
also found to be resolved in {\em all} six of our observations, with a
deconvolved mean full width at half maximum of about 0.95 mas.  This
is somewhat smaller than the figure reported by Andr\'e et al.\
(1991), but we note (i) that our observations and those of Andr\'e et
al.\ (1991) were obtained at different wavelengths; and (ii) that at
some of our epochs, the size of the emission reached 1.5 mas, whereas
at other epochs, it was smaller than 0.5 mas. At the distance of S1
(see below), 0.95 mas corresponds to about 24 \Rsun. The diameter of
S1 is expected to be about 8.5 \Rsun\ (Andr\'e et al. 1991), so its
magnetosphere appears to be on average 3 times more extended than its
photosphere.

The fact that S1 is resolved, and that its size varies from epoch to
epoch likely produces small random shifts in the photocenter of the
radio emission with an amplitude of a fraction of the size of the
emitting region. The true uncertainties on the position of S1 are,
therefore, likely to be somewhat larger than the figures quoted in
Tab.\ 1. Another factor that must be taken into account is that S1 is
known to be a member of a binary system with a separation of about 20
mas (Richichi et al.\ 1994). The companion is inferred to be about 4
times dimmer than the primary at K band, so it is likely to be
significantly less massive (Richichi et al.\ 1994). If we assume S1 to
be a 6 \Msun\ star (as suggested by its B4 spectral type), we expect
the orbital period to be about 0.7 yr, and the reflex motion of S1 to
be about 1 to 2 mas if the companion is 10 to 20 times less massive
than S1. Thus, the amplitude of the reflex motion is expected to be
larger than the formal errors on the positions of S1 listed in Tab.\
1.

\subsection{Properties of DoAr21}

The total radio flux of DoAr21 has long been known to be highly
variable (Feigelson \& Montmerle 1985). Our observations certainly
confirm this strong variability since the ratio between the highest
and the lowest measured flux exceeds 50 (Fig.\ 1). In particular, the
flux during our first two observations (10--20 mJy) is systematically
about an order of magnitude higher than that (0.4--2 mJy) at any of
the following 5 observations. Unfortunately, our time coverage is too
coarse to decide whether these first two epochs correspond to two
different flares, or to a single long-duration one.

The extreme variability of DoAr21, while at odds with the situation in
S1, is reminiscent of the case of the spectroscopic binary V773 Tau
(e.g.\ Massi et al.\ 2002). In the latter source, Massi et al.\ (2002)
showed that the variability had the same periodicity as the orbital
motion, with the radio flux being highest at periastron.
Interestingly, DoAr21 was found to be double during our second
observation\footnote{The position given in Tab.\ 1 is that of the
  brightest of the two components. The other source is offset by more
  than 5 mas from its position of the steady component expected from
  the astrometry fits presented in \S 4.3.}. This suggests that the
  same mechanism that enhances the radio emission when the two binary
  components are nearest, might be at work in both objects. The
  separation between the two components of DoAr21 in our second
  observation is about 5 mas.  This value, of course, corresponds to
  the projected separation; the actual distance between them must be
  somewhat larger. Moreover, if the mechanisms at work in DoAr21 and
  V773 Tau are similar, then DoAr21 must have been near periastron
  during our second epoch, and the orbit must be somewhat
  eccentric. As a consequence of these two effects, the semi-major
  axis of the orbit is likely to be a few times larger than the
  measured separation between the components at our second epoch,
  perhaps 10 to 15 mas. At the distance of DoAr21, this corresponds to
  1.2 to 1.8 AU. For a mass of 2.2 \Msun\ (see Sect.\ 1), the
  corresponding orbital period is 0.4 to 1.3 yr, and one would expect
  the source to oscillate with this kind of periodicity.

\subsection{Astrometry}

The absolute positions of S1 and DoAr21 (listed in columns 3 and 4 of
Tab.\ 1) were determined using a 2D Gaussian fitting procedure (task
JMFIT in AIPS). This task provides an estimate of the position errors
(also given in columns 3 and 4 of Tab.\ 1) based on the expected
theoretical astrometric precision of an interferometer (Condon
1997). Systematic errors, however, usually limit the actual precision
of VLBI astrometry to several times this theoretical value (e.g.\
Pradel et al.\ 2006, Loinard et al.\ 2007).  Moreover, we have just
seen that the extended magnetosphere of S1, and the reflex motions of
both S1 and DoAr21 are likely to produce significant shifts in the
positions of the source photocenters. While the effect of an extended
magnetosphere might be to produce a random jitter, the reflex orbital
motions ought to generate oscillations with a periodicity equal to
that of the orbital motions.  Our observations, however, are currently
insufficient to properly fit full Keplerian orbits. Instead, in the
present paper, we represent the possible systematic calibration errors
as well as the jitter due to extended magnetospheres and the
oscillations due to reflex motions, by a constant error term (the
value of which will be determined below) that we add quadratically to
the errors given in Tab.\ 1. The displacements of both S1 and DoAr21
on the celestial sphere are then modeled as a combination of their
trigonometric parallaxes ($\pi$) and their proper motions
($\mu_\alpha$ and $\mu_\delta$), assumed to be uniform and linear. The
astrometric parameters were determined using a least-square fit based
on a Singular Value Decomposition (SVD) scheme (see Loinard et al.\
2007 for details). The reference epoch was taken at the mean of each
set of observations (JD 2453757.63 $\equiv$ J2006.061 for S1, and JD
2453796.52 $\equiv$ J2006.167 for DoAr21). The best fit for S1 (Fig.\
2a) yields the following astrometric parameters:

\begin{eqnarray}
\alpha_{J2006.061} & = & \mbox{ \dechms{16}{26}{34}{174127} } ~ \pm ~ \mbox{ \mmsec{0}{000026} } \nonumber \\%
\delta_{J2006.061} & = & \mbox{ \decdms{-24}{23}{28}{44498} } ~ \pm ~ \mbox{ \msec{0}{00028} } \nonumber \\%
\mu_\alpha \cos \delta & = & -3.88 ~ \pm ~ 0.87 ~ \mbox{mas yr$^{-1}$} \nonumber \\%
\mu_\delta & = & -31.55 ~ \pm ~ 0.69 ~ \mbox{mas yr$^{-1}$} \nonumber \\%
\pi & = & 8.55 ~ \pm ~ 0.50 ~ \mbox{mas.} \nonumber
\end{eqnarray}

\noindent
For DoAr21, on the other and, we get (Fig.\ 2b):

\begin{eqnarray}
\alpha_{J2006.167} & = & \mbox{ \dechms{16}{26}{03}{018535} } ~ \pm ~ \mbox{ \mmsec{0}{000020} } \nonumber \\%
\delta_{J2006.167} & = & \mbox{ \decdms{-24}{23}{36}{35830} } ~ \pm ~ \mbox{ \msec{0}{00022} } \nonumber \\%
\mu_\alpha \cos \delta & = & -26.47 ~ \pm ~ 0.92 ~ \mbox{mas yr$^{-1}$} \nonumber \\%
\mu_\delta & = & -28.23 ~ \pm ~ 0.73 ~ \mbox{mas yr$^{-1}$} \nonumber \\%
\pi & = & 8.20 ~ \pm ~ 0.37 ~ \mbox{mas.} \nonumber
\end{eqnarray}

\noindent
To obtain a reduced $\chi^2$ of 1 in both right ascension and
declination, one must add quadratically 0.062 ms of time, and 0.67 mas
to the statistical errors of S1 listed in Tab.\ 1, and 0.053 ms of
time, and 0.57 mas to the statistical errors of DoAr21. These figures
include all the unmodeled sources of positional shifts mentioned
earlier. Interestingly, the residuals of the fit to the S1 data (inset
in Fig.\ 2a) are not random, but seem to show a $\sim$ 0.7 yr
periodicity, as expected from the reflex motions (Sect.\
4.1). Similarly, the residuals from the fit to DoAr21 seem to show a
periodicity of $\sim$ 1.2 yr (Fig.\ 2b, inset), within the range of
expected orbital periods of that system (Sect.\ 4.2). This suggests
that the errors are largely dominated by the unmodeled binarity of
both sources, and that additional observations designed to provide a
better characterization of the orbits ought to improve significantly
the precision on the trigonometric parallax determinations. 

The distance to S1 deduced from the parallax calculated above is
116.9$^{+7.2}_{-6.4}$, while the distance deduced for DoAr21 is
121.9$^{+5.8}_{-5.3}$.  The weighted mean of these two parallaxes is
8.33 $\pm$ 0.30, corresponding to a distance of
120.0$^{+4.5}_{-4.3}$. Since both S1 and DoAr21 are {\it bone fide}
members of the Ophiuchus core, this figure must represent a good
estimate of the distance to this important region of
star-formation. Note that it is in good agreement with several recent
determinations (e.g.\ de Geus et al.\ 1989, Knude \& Hog 1998,
Lombardi et al.\ ibid), but with a significantly improved relative
error of 4\%. This level of accuracy is likely to be further improved
once additional observations of S1 and DoAr21 designed to characterize
their orbital motions are available. Such observations are currently
being collected at the VLBA. A significant improvement in the distance
estimate will also be obtained once the parallax to other sources
(also currently observed at the VLBA) are measured.

\acknowledgements L.L., R.M.T, and L.F.R. acknowledge the financial
support of DGAPA, UNAM and CONACyT, M\'exico. NRAO is a facility of
the National Science Foundation operated under cooperative agreement
by Associated Universities, Inc.

{
\rotate
\begin{table*}
\caption{Observation results}
\centerline{\begin{tabular}{lcccrc}
\hline
\multicolumn{1}{c}{~~~~~~~~~~~~~~~Date~~~~~~~~~~~~~~~} & JD & $\alpha$ (J2000.0) & $\delta$ (J2000.0) & \multicolumn{1}{c}{Flux}  & Noise \\%
                                   &    & 16$^h$26$^m$ & --24$^\circ$23$'$     & \multicolumn{1}{c}{(mJy)} & (mJy beam$^{-1}$)    \\%
\hline
{\bf S1}\\%
2005 Jun 24 \dotfill & 2453545.73 & \mmsec{34}{1739533} $\pm$ \mmsec{0}{0000015} & \msec{28}{426953} $\pm$ \msec{0}{000056} & 7.03 $\pm$ 0.56 & 0.28 \\%
2005 Sep 15 \dotfill & 2453628.50 & \mmsec{34}{1736922} $\pm$ \mmsec{0}{0000020} & \msec{28}{432094} $\pm$ \msec{0}{000062} & 4.56 $\pm$ 0.47 & 0.23 \\%
2005 Dec 17 \dotfill & 2453722.25 & \mmsec{34}{1743677} $\pm$ \mmsec{0}{0000012} & \msec{28}{441493} $\pm$ \msec{0}{000044} & 4.35 $\pm$ 0.35 & 0.19 \\%
2006 Mar 15 \dotfill & 2453810.01 & \mmsec{34}{1746578} $\pm$ \mmsec{0}{0000019} & \msec{28}{451273} $\pm$ \msec{0}{000048} & 5.33 $\pm$ 0.41 & 0.17 \\%
2006 Jun 03 \dotfill & 2453889.79 & \mmsec{34}{1740172} $\pm$ \mmsec{0}{0000006} & \msec{28}{455940} $\pm$ \msec{0}{000023} & 3.29 $\pm$ 0.13 & 0.07 \\%
2006 Aug 22 \dotfill & 2453969.54 & \mmsec{34}{1732962} $\pm$ \mmsec{0}{0000012} & \msec{28}{462601} $\pm$ \msec{0}{000050} & 4.35 $\pm$ 0.22 & 0.09 \\%
\hline
{\bf DoAr21}\\%
2005 Sep 08 \dotfill & 2453621.52 & \mmsec{03}{0189304} $\pm$ \mmsec{0}{0000065} & \msec{36}{343394} $\pm$ \msec{0}{00013} & 11.78 $\pm$ 1.41 & 0.35 \\%
2005 Nov 16 \dotfill & 2453691.33 & \mmsec{03}{0191097} $\pm$ \mmsec{0}{0000023} & \msec{36}{344504} $\pm$ \msec{0}{00005} & 20.34 $\pm$ 1.42 & 0.55 \\%
2006 Jan 08 \dotfill & 2453744.19 & \mmsec{03}{0191069} $\pm$ \mmsec{0}{0000059} & \msec{36}{355803} $\pm$ \msec{0}{00023} &  0.39 $\pm$ 0.12 & 0.05 \\%
2006 Jan 19 \dotfill & 2453755.16 & \mmsec{03}{0191795} $\pm$ \mmsec{0}{0000028} & \msec{36}{355677} $\pm$ \msec{0}{00013} &  0.97 $\pm$ 0.19 & 0.11 \\%
2006 Mar 28 \dotfill & 2453822.97 & \mmsec{03}{0189625} $\pm$ \mmsec{0}{0000070} & \msec{36}{361924} $\pm$ \msec{0}{00020} &  1.49 $\pm$ 0.28 & 0.13 \\%
2006 Jun 04 \dotfill & 2453890.78 & \mmsec{03}{0182041} $\pm$ \mmsec{0}{0000019} & \msec{36}{363763} $\pm$ \msec{0}{00010} &  1.92 $\pm$ 0.23 & 0.11 \\%
2006 Aug 24 \dotfill & 2453971.53 & \mmsec{03}{0169857} $\pm$ \mmsec{0}{0000037} & \msec{36}{369957} $\pm$ \msec{0}{00016} &  1.45 $\pm$ 0.32 & 0.16 \\%
\hline
\\%
\end{tabular}}
\end{table*}}

\clearpage

\begin{figure}[!b]
\centerline{\includegraphics[height=0.4\textwidth,angle=-90]{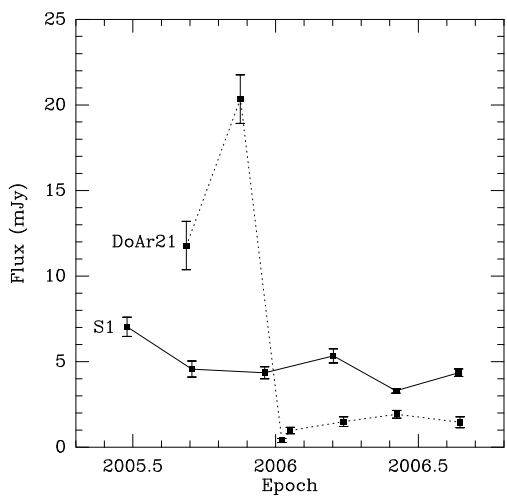}}
\caption{Radio flux of S1 (full line) and DoAr21 (dotted line) as a function 
of time.}  
\end{figure}

\begin{figure*}[!t]
\centerline{\includegraphics[height=1\textwidth,angle=-90]{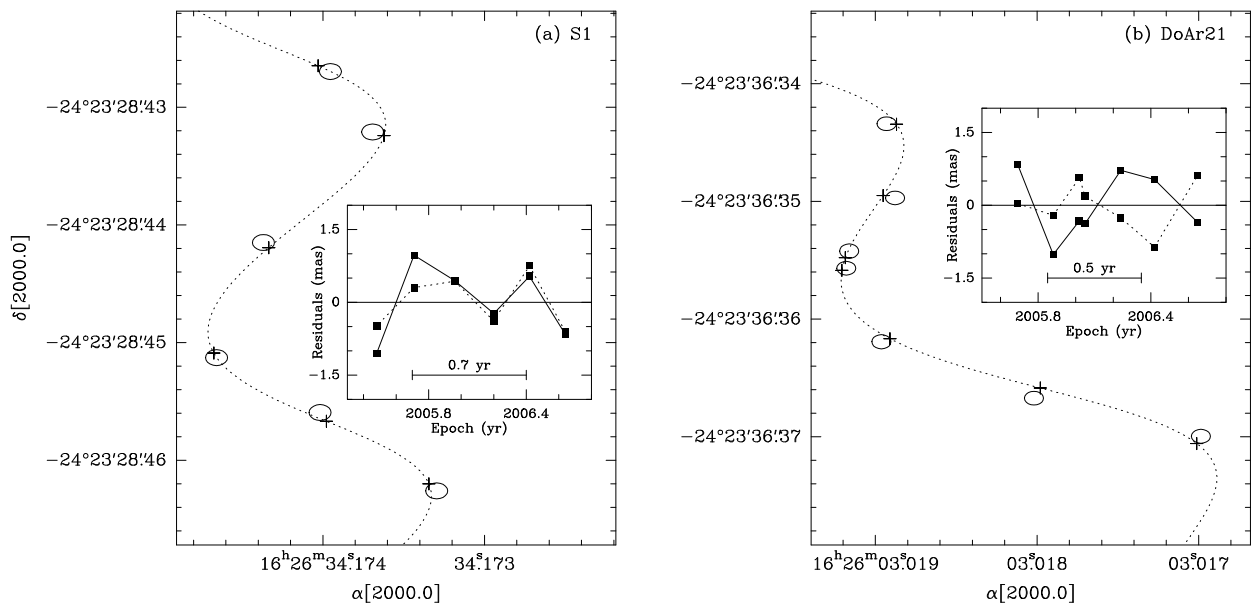}}
\caption{Measured positions and best fit for (a) S1, and (b) DoAr21.
  The observed positions are shown as ellipses, the size of which
  represents the magnitude of the errors. The positions at each epoch
  expected from the best fits are shown as $+$ signs. The insets show
  the residuals (fit-observation) in right ascension (full line) and
  declination (dotted line).}
\end{figure*}


\begin{thebibliography}{}

\bibitem[Andre et al.(1987)]{1987AJ.....93.1182A} Andre, P., Montmerle, T., 
\& Feigelson, E.~D.\ 1987, \aj, 93, 1182 

\bibitem[Andre et al.(1988)]{1988ApJ...335..940A} Andre, P., Montmerle, T., 
Feigelson, E.~D., Stine, P.~C., \& Klein, K.-L.\ 1988, \apj, 335, 940 

\bibitem[Andre et al.(1990)]{1990A&A...240..321A} Andre, P., Montmerle, T., 
Feigelson, E.~D., \& Steppe, H.\ 1990, \aap, 240, 321 

\bibitem[Andre et al.(1991)]{1991ApJ...376..630A} Andre, P., Phillips, 
R.~B., Lestrade, J.-F., \& Klein, K.-L.\ 1991, \apj, 376, 630 

\bibitem[Andre et al.(1992)]{1992ApJ...401..667A} Andre, P., Deeney, B.~D., 
Phillips, R.~B., \& Lestrade, J.-F.\ 1992, \apj, 401, 667 

\bibitem[Beasley et al.(2002)]{2002ApJS..141...13B} Beasley, A.~J., Gordon, 
D., Peck, A.~B., Petrov, L., MacMillan, D.~S., Fomalont, E.~B., \& Ma, C.\ 
2002, \apjs, 141, 13 

\bibitem[Chini(1981)]{1981A&A....99..346C} Chini, R.\ 1981, \aap, 99, 346 

\bibitem[Condon(1997)]{1997PASP..109..166C} Condon, J.~J.\ 1997, \pasp, 
109, 166 

\bibitem[Duch{\^e}ne et al.(2004)]{2004A&A...427..651D} Duch{\^e}ne, G., 
Bouvier, J., Bontemps, S., Andr{\'e}, P., \& Motte, F.\ 2004, \aap, 427, 
651 

\bibitem[Fazio et al.(1976)]{1976ApJ...206L.165F} Fazio, G.~G., Low, F.~J., 
Wright, E.~L., \& Zeilik, M., II 1976, \apjl, 206, L165 

\bibitem[Feigelson \& Montmerle(1985)]{1985ApJ...289L..19F} Feigelson, 
E.~D., \& Montmerle, T.\ 1985, \apjl, 289, L19 

\bibitem[]{884}
Feigelson, E.D., \& Montmerle, T., 1999, ARAA, 37, 363

\bibitem[Gagn{\'e} et al.(2004)]{2004ApJ...613..393G} Gagn{\'e}, M., 
Skinner, S.~L., \& Daniel, K.~J.\ 2004, \apj, 613, 393 

\bibitem[de Geus et al.(1989)]{1989A&A...216...44D} de Geus, E.~J., de 
Zeeuw, P.~T., \& Lub, J.\ 1989, \aap, 216, 44 

\bibitem[Grasdalen et al.(1973)]{1973ApJ...184L..53G} Grasdalen, G.~L., 
Strom, K.~M., \& Strom, S.~E.\ 1973, \apjl, 184, L53 

\bibitem[Greisen(2003)]{Gre03} Greisen, E.W.\ 2003, in
   Information Handling in Astronomy -- Historical Vistas, ed.\ A.\ Heck
   (Dordrecht: Kluwer Academic Publishers), 109

\bibitem[Haisch et al.(2002)]{2002AJ....124.2841H} Haisch, K.~E., Jr., 
Barsony, M., Greene, T.~P., \& Ressler, M.~E.\ 2002, \aj, 124, 2841 

\bibitem[Johnstone et al.(2004)]{2004ApJ...611L..45J} Johnstone, D., Di 
Francesco, J., \& Kirk, H.\ 2004, \apjl, 611, L45 

\bibitem[Knude \& Hog(1998)]{1998A&A...338..897K} Knude, J., \& Hog, E.\ 
1998, \aap, 338, 897 

\bibitem[Lada \& Lada(2003)]{2003ARA&A..41...57L} Lada, C.~J., \& Lada, 
E.~A.\ 2003, \araa, 41, 57 

\bibitem[Leous et al.(1991)]{1991ApJ...379..683L} Leous, J.~A., Feigelson, 
E.~D., Andre, P., \& Montmerle, T.\ 1991, \apj, 379, 683 

\bibitem[Loinard et al.(2007)]{2007ApJ...671..546L} Loinard, L., Torres, 
R.~M., Mioduszewski, A.~J., Rodr{\'{\i}}guez, L.~F., 
Gonz{\'a}lez-L{\'o}pezlira, R.~A., Lachaume, R., V{\'a}zquez, V., \& 
Gonz{\'a}lez, E.\ 2007, \apj, 671, 546 

\bibitem[Loren et al.(1990)]{1990ApJ...365..269L} Loren, R.~B., Wootten, 
A., \& Wilking, B.~A.\ 1990, \apj, 365, 269 

\bibitem[Mamajek(2007)]{2007arXiv0709.0505M} Mamajek, E.~E.\ 2007, ArXiv 
e-prints, 709, arXiv:0709.0505 

\bibitem[Massi et al.(2002)]{2002A&A...382..152M} Massi, M., Menten, K., \& 
Neidh{\"o}fer, J.\ 2002, \aap, 382, 152 

\bibitem[Montmerle et al.(1983)]{1983ApJ...269..182M} Montmerle, T., 
Koch-Miramond, L., Falgarone, E., \& Grindlay, J.~E.\ 1983, \apj, 269, 182 

\bibitem[Motte et al.(1998)]{1998A&A...336..150M} Motte, F., Andre, P., \& 
Neri, R.\ 1998, \aap, 336, 150 

\bibitem[Ozawa et al.(2005)]{2005A&A...438..661O} Ozawa, H., Grosso, N., \& 
Montmerle, T.\ 2005, \aap, 438, 661 

\bibitem[Padgett et al.(2007)]{2007arXiv0709.3492P} Padgett, D.~L., et al.\ 
2007, ArXiv e-prints, 709, arXiv:0709.3492 

\bibitem[Pradel et al.(2006)]{2006A&A...452.1099P} Pradel, N., Charlot, P., 
\& Lestrade, J.-F.\ 2006, \aap, 452, 1099 

\bibitem[Richichi et al.(1994)]{1994A&A...287..145R} Richichi, A., Leinert, 
C., Jameson, R., \& Zinnecker, H.\ 1994, \aap, 287, 145 

\bibitem[Torres et al.(2007)]{2007ApJ...671.1813T} Torres, R.~M., Loinard, 
L., Mioduszewski, A.~J., \& Rodr{\'{\i}}guez, L.~F.\ 2007, \apj, 671, 1813 

\end{thebibliography}
\end{document}